\documentclass[sigconf]{acmart}

\usepackage{booktabs} % For formal tables
\usepackage{tabulary}
\usepackage{microtype}
\usepackage{multirow}
\usepackage{amsmath}
\hypersetup{draft}

\renewcommand{\paragraph}[1]{
\noindent \textbf{#1.}
}

\usepackage{xelatexemoji}
\renewcommand{\xelatexemojipath}[1]{emoji/#1.png}

\copyrightyear{2020}
\acmYear{2020}
\setcopyright{acmlicensed}\acmConference[WebSci '20]{12th ACM Conference on Web Science}{July 6--10, 2020}{Southampton, United Kingdom}
\acmBooktitle{12th ACM Conference on Web Science (WebSci '20), July 6--10, 2020, Southampton, United Kingdom}
\acmPrice{15.00}
\acmDOI{10.1145/3394231.3397907}
\acmISBN{978-1-4503-7989-2/20/07}

\renewcommand\footnotetextcopyrightpermission[1]{\footnotetext{This is the author's version of the document. The citation and DOI are available at the ACM Reference Format note displayed above.}} % removes footnote with conference information in first column
\begin{document}
\title{Every Colour You Are:\\ Stance Prediction and Turnaround in Controversial Issues}

\author{Eduardo Graells-Garrido}
\affiliation{%
  \institution{Barcelona Supercomputing Center}
  \city{Barcelona}
  \country{Spain}
}
\email{eduardo.graells@bsc.es}
\additionalaffiliation{%
  \institution{Data Science Institute, Universidad del Desarrollo}
  \city{Santiago}
  \country{Chile}
%}
}

\author{Ricardo Baeza-Yates}
\affiliation{%
  \institution{Northeastern University at SV}
  \state{California}
  \country{USA}
}
\email{rbaeza@acm.org}

\author{Mounia Lalmas}
\affiliation{%
  \institution{Spotify}
  \streetaddress{}
  \city{London}
  \country{UK}}
\email{mounia@acm.org}

\begin{abstract}
Web platforms have allowed political manifestation and debate for decades. Technology changes have brought new opportunities for expression, and the availability of longitudinal data of these debates entice new questions regarding who participates, and who updates their opinion.
The aim of this work is to provide a methodology to measure these phenomena, and to test this methodology on a specific topic, abortion, as observed on one of the most popular micro-blogging platforms. 
To do so, we followed the discussion on Twitter about abortion in two Spanish-speaking countries from 2015 to 2018. 
Our main insights are two fold. 
On the one hand, people adopted new technologies to express their stances, particularly colored variations of heart emojis (\xelatexemoji{1f49a} \& \xelatexemoji{1f499}) in a way that mirrored physical manifestations on abortion.
On the other hand, even on issues with strong opinions, opinions can change, and these changes show differences in demographic groups.
These findings imply that debate on the Web embraces new ways of stance adherence, and that changes of opinion can be measured and characterized. 
\end{abstract}



\maketitle
\title{Every Colour You Are: Stance Prediction and Turnaround in Controversial Issues}


\sloppy
\section{Introduction}

The Web is an important medium to exchange points of view. Its several platforms have connected people and allowed manifestation, organization, and access to information. 
Longitudinal studies covering political debates exist~\cite{garimella2017long,garimella2017effect}, although they have not looked at who discusses what, and with whom.
As our case study, we focus on the debate about \emph{abortion}.
Abortion is a hard topic to talk about, as it is not only about political stances, but also about deep private matters~\cite{sanger2016talking}.
Debates on micro-blogging platforms on this topic have been studied before, describing how the different stances relate and are discussed~\cite{yardi2010dynamic,graells2015finding,zhang2016gender,sharma2017analyzing}. 

However, the aforementioned work has focused on a single country (USA), with English as main language, and textual content as main unit of analysis. 
In contrast, here we study the Twitter debate on abortion in Argentina and Chile, two neighboring Spanish-speaking countries in South America, during 2015--2018.
Although both countries share many cultural similarities, they have several differences in terms of abortion legislation and Twitter population~\cite{graells2020representativeness}. 
Chile is a country that, until the approval of its current abortion law, had one of the most severe abortion laws in the world~\cite{shepard2007abortion}. Its current abortion bill was sent to congress in January 2015 and approved as law in September 2017, eliciting discussion on social media until today.
Argentina does not have an abortion law, although its current Sanitary Code from 1921 allows abortion on two grounds: endangerment of life and pregnancy from sexual violence. 
In 2018, a free abortion law was proposed in the Argentinian Congress, but was rejected after two months of legislation~\cite{BOOTH2018e21}.
What characterizes this context is how physical manifestations and protests have connected activist movements from several countries. In the last twenty years, abortion activists in Argentina have expressed themselves through green and purple kerchiefs~\cite{sutton2020abortion}. The movement, called \emph{green wave}, has influenced how abortion activism was (is) carried out in Latin America~\cite{greenwave}.

With this context in mind, we aim to better understand how people take a stance on debates around abortion specifically, but also in controversial issues in general. More precisely, we look into how these on-going movements have characteristics that go beyond written language, and how people react to on-going legislation while these movements are active.
We aim to answer two research questions: 
RQ1) \emph{How people make use of new technologies to express their positions on controversial issues?}; and 
RQ2) \emph{Which demographic and profile factors characterize opinion change?}

To answer these questions, we designed a methodology to study opinions expressed on micro-blogging platforms. We detail how we label demographic attributes and stances in profiles taking advantage of information already available on user profiles.
Then, we describe methods that allowed us to answer our research questions, by means of inspecting how a stance classifier predicts stance, and by the linear regression over a metric of stance turnaround.

The main contribution of this work are the novel insights regarding the abortion discussion derived from applying our methodology. 
Regarding RQ1, we observed that people adopted a new technology to express their stance: emojis, {\textit i.e.}, pictures represented as encoded characters in text, part of mainstream text-input user interfaces. Particularly, heart colored emojis (\xelatexemoji{1f49a} \& \xelatexemoji{1f499}) are strong predictors of stance, both in tweet content and user profiles. %\footnote{Both emojis were included in the Unicode standard in 2010 (version 6.0.0), and have been progressively included in user interfaces since then.}
Regarding RQ2, we observed that turnarounds in abortion stance can be inferred, identifying how demographic and profile characteristics explain the variations in stance after important events on the issue, such as legislative actions in Congress. 

In conclusion, this paper showcases how the analysis of a longitudinal discussion can support the analysis of socially relevant phenomena: how people express their stances, mirroring physical manifestations, and what are the characteristics of people that change their views in time, enabling the measurement of how different events impact stance of people. In an ever-changing world, embracing Web platforms brings also new ways of expression.

\section{Related Work}
\label{relatedwork}

Twitter has been a platform that has enabled the study of controversial discussion at scale~\cite{garimella2018quantifying}. 
In relation to abortion, different perspectives have been studied: how people from each stance interacts with others~\cite{yardi2010dynamic}; the linguistic characteristics of ideological discourse~\cite{sharma2017analyzing}; and the spread of anti-abortion policy~\cite{zhang2016gender}.
To characterize stances on these types of controversial issues, stances must be predicted, as they are not always explicit. %, {\it e.g.},~self-reported. 
Two types of approaches are common. 
Stance can be predicted using network interactions, based on the assumption that like-minded people are more likely to interact~\cite{garimella2018political,garimella2018quantifying}.
In addition, lexical analysis have shown to allow predicting stance as vocabulary within stances tend to have strongly associated words~\cite{conover2011predicting,lu2015biaswatch}. 
However, stance prediction is not a fully solved problem. For instance, participating in the debate and taking a stance are two actions that are assumed to be the equivalent, but are not~\cite{zhang2019stances}. 
In our work we apply a mixed approach, and define an \emph{undisclosed} stance to account for participation in the debate without disclosing stance. %, which we utilize through the classifier prediction probabilities.

In the lexical approach, a common feature is the usage of hashtags in micro-post content ({\textit e.g.}, \#freeabortion, \#notoabortion). Hash\-tags are a form of expression native to the Web that is increasingly associated with physical manifestations of political debates and protests~\cite{steinert2015online}. 
We believe that current-generation web-technologies, such as emojis, a focus of this paper, are missing in these studies. Doing so poses challenges, for instance, as there are cultural differences in emoji usage~\cite{barbieri2016cosmopolitan}, and there can be several interpretations of the same emoji~\cite{miller2016blissfully}. Nevertheless, emojis are so popular that people constantly request new ones~\cite{feng2019world}. In addition, their usage patterns allow to predict user characteristics, such as gender~\cite{chen2018through}.
In our context, emoji has been analyzed in political discussion on social networks~\cite{liebeskind2019emoji}, but only with respect to what representation is best for predicting the emoji to be used in text, regardless of its actual meaning, {\em i.e.}, the work would be similar in a non-political discussion.
Although they are typically understood as emotional or sentiment cues~\cite{felbo2017using}, we find that emojis are a powerful predictor of abortion stance in Argentina and Chile, not only in micro-post content, but also in profile elements---mirroring how colored kerchiefs are used in physical manifestations~\cite{sutton2020abortion,greenwave}. To the extent of our knowledge, this association between stance on controversial issues and emojis has not been described before.

Longitudinal studies on political debate on Twitter exist. 
Eight years of debate on Twitter in the USA showed that polarization gradually increased in time, in terms of people supporting democrat/republican politicians~\cite{garimella2017long}. Five years of debate on the same country provided insights on how phy\-si\-cal-world events had impact on virtual discussion~\cite{garimella2017effect}.
On a smaller scale, four months of Twitter debate on judicial decisions on same-sex marriage showed that Supreme Court decisions polarized the public and generated emotional shifts in public opinion~\cite{clark2018using}.
While our motivation is similar to longitudinal works such as~\citet{garimella2017effect}, our analysis differs in focus.
Our attention is not on how polarization evolves in time, instead, we identify users who have changed their views before and after an important event for the discussion, and find which demographics and profile characteristics relate to stance turnaround (or the absence of it). 

\begin{figure*}[t]
\centering
\includegraphics[width=0.90\linewidth]{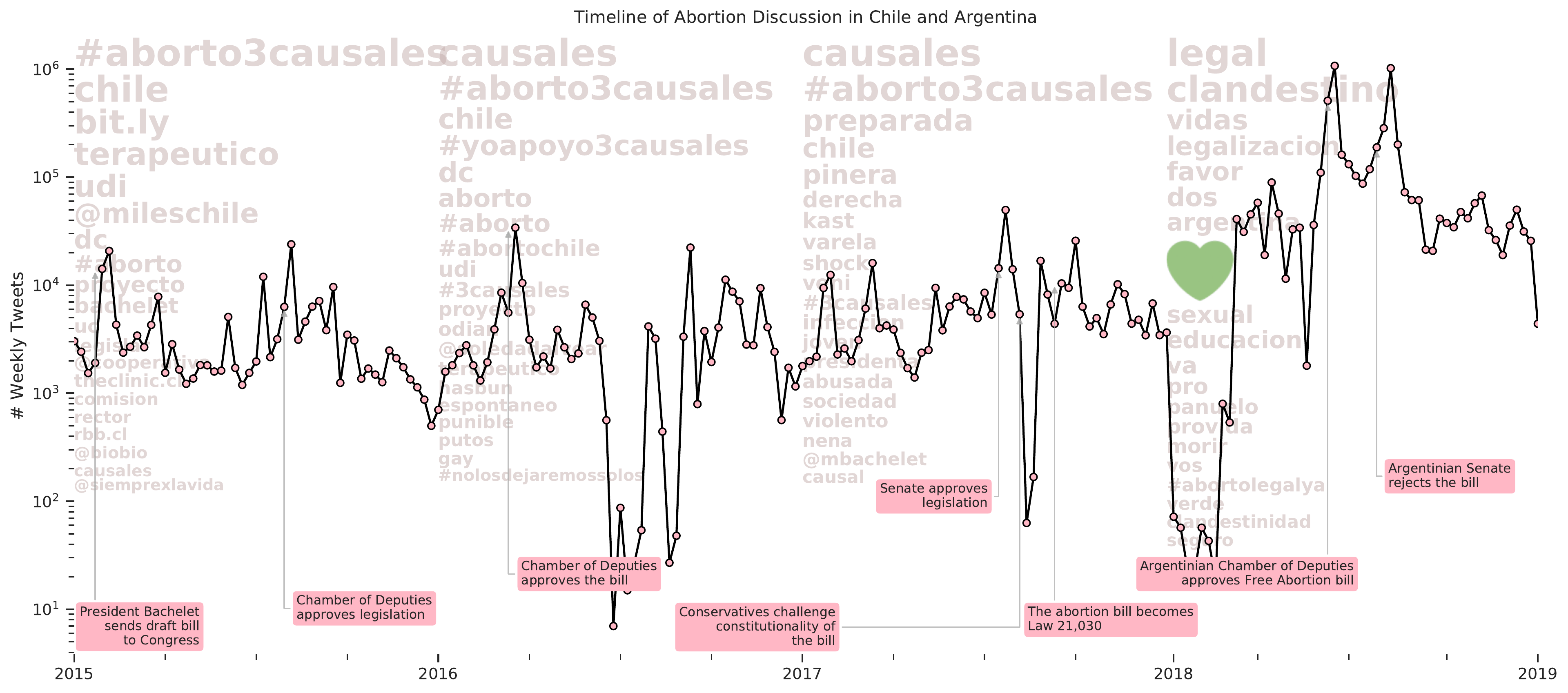}
\caption{Weekly tweet volume in the period of study. Each box represents a legislation event. Events in 2015--2017 happened in the Chilean Congress. Events in 2018 happened in the Argentinian Congress. Words represent the most relevant terms for every year of discussion. Some weeks presented crawling interruptions (in 07/2016, 09/2016, 12/2017, 01/2018).}
\label{fig:timeline}
\end{figure*}

\section{Dataset}
\label{materials}

We describe the dataset used to study the abortion legislation debate on Twitter, the case study in this work. The analysis has a temporal coverage of four years of Twitter debate (2015--2018), a geographical coverage of two countries (Argentina and Chile), and a political coverage of two abortion laws (one approved, one rejected). 
Chronologically, the debate started with the abortion bill proposed in 2015 by Chilean former President Bache\-let, and approved as law in 2017.
In Argentina, the legislative debate was held in 2018, during President Macri's period. 

On Twitter, users have a profile and publish micro-posts, usually with limited number of characters allowed (currently 280). 
Each micro-post (a \emph{tweet}) may contain multimedia items, hashtags (or topic indicators, \emph{e.g.}, \#abortion), mentions of others users, and links to websites. 
A micro-post may also be published again by someone else than its original author (in Twitter, this is known as \emph{retweeting}), cited or quoted, and replied to. 

User profiles in Twitter contain the following features: a \emph{screen-name} or alias, a full name (which may not be validated), an optional location in free text form (eventually fictional~\cite{hecht2011tweets}), an optional self-description or biography, an optional URL, the number of published micro-posts, the number of followers or subscribers, and the number of friends or subscriptions to other profiles. 

We crawled tweets using the Twitter Streaming API between January 1, 2015, and December 31, 2018. 
The query parameters were keywords related to abortion, composed into a query using the OR operator, and applied to the tweet content.
The keyword set included general abortion vocabulary (\emph{e.g.}, aborto(s), tenses of \textit{to abort} in Spanish), hashtags, both general (\emph{e.g.}, \#aborto3causales --abortion three grounds--, \#noalaborto --no to abortion--)\footnote{In Chile, the law allows abortion under three grounds: ``endangerment of a woman's life; embryonic anomaly or  malformation incompatible with life; and  pregnancy arising from sexual violence''~\cite{montero2018critical}.} and contextual (\emph{e.g.}, \#marchaabortolegal --protest for legal abortion--), mentions to accounts involved in the debate (\emph{e.g.}, public health institutions, NGOs), and phrases (\emph{e.g.}, ``pregnancy interruption'').

Initially, the dataset contained 31.4M tweets from 1.8M users. 
This dataset was pre-processed and filtered to ensure that we analyzed discussion and debate.
Firstly, we identified users who self-reported their gender (male, female) and country (Argentina, Chile) on their profiles. We then proceeded to propagate these labels to the rest of the dataset using a classifier (Section \ref{methods} details this process). Only those profiles with valid gender and location attributes (either self-reported or predicted with high confidence) were kept.
Secondly, we filtered out users that did not belong to the largest connected component (LCC) of the discussion network (comprised of retweets, mentions, replies, and quotes). The LCC contained 84\% of the nodes in the network, and the second LCC had less than 0.01\% of nodes.
As a result, the final dataset was comprised by 6M tweets from 663K users. 

Figure~\ref{fig:timeline} shows the weekly volume of tweets and yearly-relevant terms.
We observe several peaks, most of them occurring around legislative events. 
Word relevance was estimated using Log-Odds Ratio with Uninformative Dirichlet Prior~\cite{monroe2008fightin}, which weights the frequency of words in a similar way to TF-IDF, but without over-weighting low frequency features. 
Relevant words include topic related ones, such as \emph{clandestino} (\emph{clandestine}, 2018), \emph{terapéutico} (\emph{therapeutic}, 2015), \emph{causales} (\emph{grounds}, 2017); political parties (\emph{DC} and \emph{UDI} from Chile, in 2015 and 2016); and politicians (\emph{@mbachelet} and \emph{Kast} from Chile, 2017). 
The weight of \xelatexemoji{1f49a} in 2018, and related words to it (\emph{pañuelo} --\emph{kerchief}--, \emph{verde} --\emph{green}--), hints that such graphical elements may have some importance in our analysis. 

\null

We analyze this abortion debate to contribute insights on two aspects: the expressiveness of stance in web-based technology and the analysis of turnarounds. In the next section we describe the methodology applied to this dataset.

\begin{figure*}[t]
\centering
\includegraphics[width=0.95\linewidth]{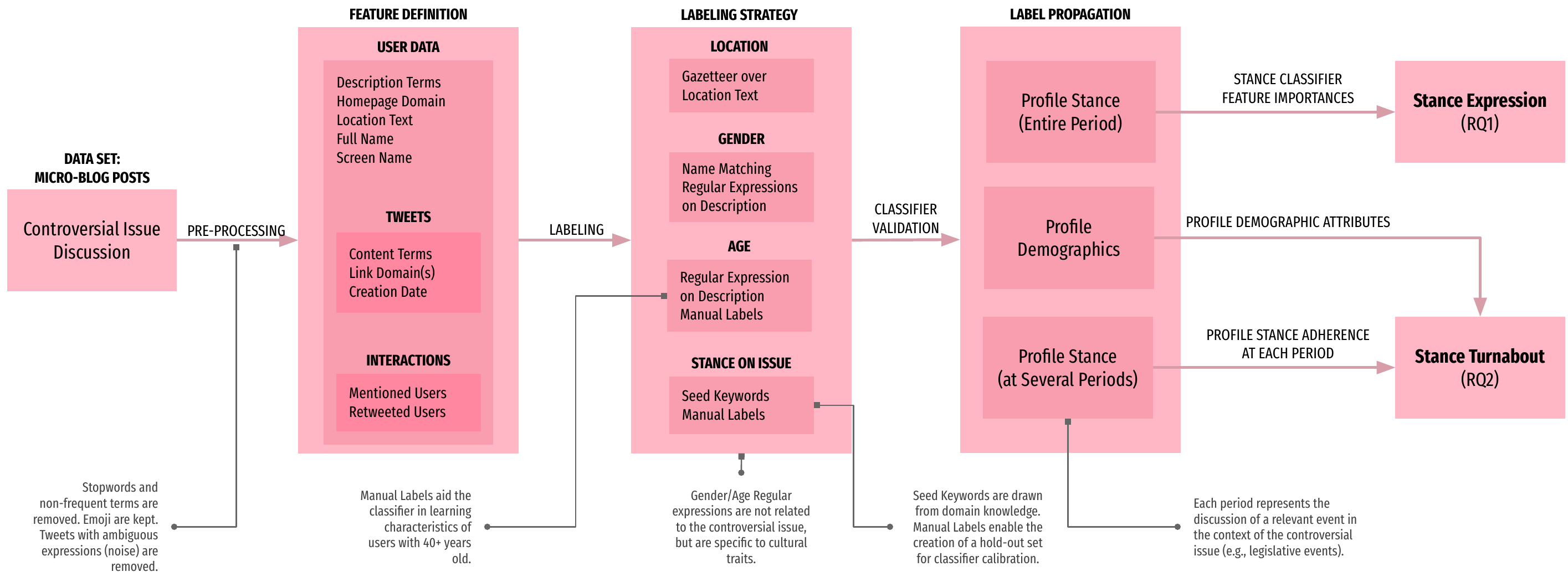}
\caption{Schematic diagram of the proposed methodology.}
\label{fig:methodology}
\end{figure*}

\section{Methodology}
\label{methods}
This section details our proposed methodology to analyze discussions on controversial issues. 
The methodology is defined through the following steps: \emph{data pre-processing}, \emph{user profile labeling with demographic and stance attributes}, \emph{classifier validation and application}, and finally how we \emph{use the user profiles and the outcomes of the classifier} to answer our research questions. A schematic diagram of the methodology is shown on Figure~\ref{fig:methodology}.

\subsection{Data Pre-Processing}

At this stage, the dataset of micro-posts contains potentially relevant content related to the controversial issue under study. To ensure relevancy, a typical pre-processing pipeline includes filtering based on rules to  discard content that contains keywords associated with the discussion, but are used in a different context.\footnote{Expressions regarding abortion as insult are common in Argentina, as in ``you were aborted by a whale.'' It was an enormous effort to account for these informalities.}

The purpose here is to build a feature matrix from relevant content, where each user is a row and each feature is a column. The features include user data such as  profile, micro-posts published, and interactions with other users.
The feature matrix itself is a horizontal concatenation of multiple matrices, defined as follows: 
\begin{enumerate}
    \item A user-term matrix, where a cell $(i, j)$ contains the number of times user $i$ has used the term $j$ in his/her/their tweets. Terms include words, hashtags, user names, URLs, and emojis. We only consider terms that occur at least 50 times, and we discard stopwords.
    \item A second user-term matrix, where a cell $(i, j)$ contains the number of times user $i$ has used the term $j$ in his/her/their biographical description. We only consider terms that occur at least 10 times. We built a custom lexicon of words with semantic categories, which are also considered in this matrix. The lexicon includes categories such as social media ({\em e.g.}, \emph{Ig} or \emph{Snap}), profession words, celebrity accounts, etc.
    \item A matrix of one-hot encoded features. These features include profile meta-data such as the domain of the user home page, a time-zone (if available), the number of emojis in the profile description, and the use of each emoji in the account's reported name. 
    \item Two adjacency matrices. One of retweets, where a cell $(i, j)$ contains the number of times user $i$ has retweeted user $j$; and one of mentions, replies, and quotes, where a cell  $(i, j)$ contains the number of times user $i$ has written tweets directed at user $j$. %These matrices contain only users in our dataset. 
\end{enumerate}

This feature matrix is used in the following two steps, user profile labeling, and classifier validation and application.

\subsection{User Profile Labeling}
To answer our research questions, we need to know various characteristics of user profiles. We predict them using a classifier, but first we need labeled data. This step in our methodology takes care of generating labeled subsets of users for each needed profile characteristic. The overall approach is simple, but non trivial: we rely on self-reported information, both in explicit and implicit disclosures of the relevant attributes.

\subsubsection{Location (binary)}
To label a user location, we check their self-reported location name against a manually built gazetteer for all locations within the countries of interest.

\subsubsection{Gender (binary)}
To label a user gender, we first check their first name in the reported full name, by checking from a list of known names. For users without an identifiable name, we check their biographies for typical expressions disclosing gender (\emph{e.g.}, ``Mother of two kids''). This list of expressions is built manually, as it is dependant on the main language of the dataset (but not on the specific issue under analysis).\footnote{Spanish is a gendered language. In contrast to English, neutral pronouns do not exist.}

\subsubsection{Age (cohorts)}
To label age we match common phrases in biographies that contained age or date of birth (\emph{e.g.}, ``25 years old''~\cite{sloan2015tweets} in Spanish). Instead of focusing on the exact age, users are grouped into four age cohorts ($<$ 18, 18--29, 30--39, $\geq$ 40). Identifying people in the fourth cohort is arguably harder, as they are less likely to report their age (based on our experience). Thus, manual labels of these profiles may be required.

\subsubsection{Stance on a Controversial Issue (abortion, binary)}
To label stance we follow two strategies. The first one is similar to labeling demographic attributes, with the exception that instead of patterns in phrases, we look for seed patterns and keywords associated with each stance. The second one is to build a set of manually labeled users that do not comply with these patterns.

Before applying seed patterns and keywords, we need to define the actual stances under study.
In abortion there are two main stances, colloquially denoted \emph{pro-life} and \emph{pro-choice}. 
Although commonly used, these terms are semantically overloaded, \emph{i.e.}, they carry an implicit false leaning on behalf of the opposite stance~\cite{greasley2017arguments}. 
Therefore, we adopt these two stances: \emph{opposition} (instead of pro-life), and \emph{defense} (instead of pro-choice). 
We use domain knowledge from our previous work analyzing the abortion discussion~\cite{graells2020representativeness} to build a list of seed patterns and keywords. Table~\ref{table:stance_validation} shows a subset of the seed patterns and keywords used to label users to each stance in Argentina and Chile, including words, expressions, hashtags, activist accounts, and campaign accounts associated with each stance (the table includes translations and explanations). 
Given that keyword usage is not exclusive to each stance (for instance, due to hashtag hijacking~\cite{hadgu2013political}), we only label users that match these patterns for one stance but not for the other.

\begin{table}
    \centering
    \footnotesize
    \caption{Seed patterns for each abortion stance.}
    \begin{tabulary}{\linewidth}{l|L|L}
    \toprule
    Stance & Patterns in Biography & Patterns in Tweets \\
    \midrule
    \emph{defense} & \#abortolegal (\textit{\#legalabortion}), \#abortolibre (\textit{\#freeabortion}), \#abortoseguro (\textit{\#safeabortion}), \#abortogratuito (\textit{\#freeabortion}), feminista (\textit{\#feminist}), a favor del aborto (\textit{supports abortion}), \#proeleccion (\textit{\#prochoice}), \#prochoice & @abortolegalcl (activist account), \#nobastantrescausales (\textit{\#threegroundsarenotenough}), \#seraley (\textit{\#itwillbelaw}), @CampAbortoLegal (campaign) \\
    \midrule
    \emph{opposition} & @siemprexlavida (\textit{@alwaysxlife}), derecho a la vida (\textit{right to live}), \#antiaborto (\textit{\#antiabortion}), contrario/a al aborto (\textit{against abortion}), \#stopaborto, las dos vidas (\textit{the two lives}), cristiano/a (\textit{christian}), \#salvemoslasdosvidas (\textit{\#savethetwolives}), aborto no es la solución (\textit{abortion is not the solution}), \#siempreporlavida (\textit{\#alwaysforlife}), \#porlasdosvidas (\textit{\#forthetwolives}), \#profamilia (\textit{\#profamily}), \#provida (\textit{\#prolife}), \#noalaborto (\textit{\#notoabortion}) & \#salvemoslasdosvidas (\textit{\#letssavethetwolives}), \#sialavida (\textit{\#yestolife}), \#abortolegalno (\textit{\#notolegalabortion}), \#noalaborto (\textit{\#notoabortion}), \#noesley (\textit{\#itsnotlaw}), @mmreivindica (activist account), @noalaborto\_arg (campaign) \\
    \bottomrule
    \end{tabulary}
    \label{table:stance_validation}
    \end{table}

\null 

As result, we have a labeled subset of the dataset with demographic and stance labels. Next, we describe how to propagate these labels to the rest of the dataset.

\subsection{Classifier Validation and Application}

We follow a bootstrapped approach to predict profile characteristics for all users: we learn using the labels obtained from the previous step, and then propagate them to the rest of the dataset~\cite{pennacchiotti2011machine}. 

\subsubsection{Classifier Training and Validation}
For each profile characteristic we want to predict, we train a XGBoost (XGB) classifier~\cite{chen2016xgboost}. XGB is a gradient boosting algorithm based on decision trees. It works by training a number of weak learners using randomly built small subsets of the data, and then returning the fraction of learners that predict each possible category.%
\footnote{The parameters for each classifier are as follows: 300 \emph{estimators} (the number of trees or weak learners), a \emph{learning rate} of 0.1 (how much the feature weights decrease after each iteration to avoid over-fitting), and a \emph{max delta step} of 1 (the maximum change in a tree leaf from one iteration to the next).}
To avoid over-fitting we train with early stopping, by using a validation set of 20\% of the training observations.
We also remove columns from the feature matrix that were used for labeling. This includes numbers with two or four digits, used in age cohort labeling, and seed keywords for each stance, as they perfectly separate users from both groups---and our aim is to predict a stance for users that do not use these terms in their content.
Finally, to validate the predictive performance of each classifier, we measure precision and recall using 5-fold cross-validation.

\subsubsection{Predicting Demographic Characteristics}
In tree-based models, the classification output is the proportion or weak learners that vote for each stance in a prediction, which we interpret as the prediction confidence of a given profile characteristic for a specific user. 
We only accept classification outcomes above the following thresholds: 0.7 for gender and location, and 0.65 for age (selected manually through experimentation).

\subsubsection{Stance Adherence Probabilities}
Since in abortion there are two stances, the prediction confidence of one stance is symmetrical with respect to the other, which allows us to work using one stance as unit of analysis without losing generality. However, the complexities of the issue under analysis pose two challenges.
First, using seed keywords to label user stance means that only users with extreme positions on the stance spectrum are labeled. Hence, the XGB classifier makes the implicit assumption that all users have extreme views on the issue. 
This may not be realistic, as not all people has their views positioned in those extremes, and some people may have an undecided position~\cite{zhang2019stances}.
Second, it is not clear if the proportion of stances in the labeled set of users is the same as in the non-labeled set~\cite{card2018importance}. For instance, a prediction confidence of 0.5 may not imply an undecided profile, because these values are not real probabilities.

We address both challenges by converting stance prediction confidence into probabilities by calibrating the classifier~\cite{card2018importance}. For calibration to work effectively, the dataset used for this purpose should be disjoint from the one used for training. Here we use the set of manually labeled users with stance from the previous step to calibrate the XGB prediction outcomes with Platt's calibration method~\cite{niculescu2005predicting}, converting them into stance adherence probabilities.
This allows us to classify users into three classes: \emph{opposition} (probability of \emph{defense} label $p(\text{defense})$ within $[0, 0.4[$), \emph{undisclosed} ($p(\text{defense})$ within $(0.4, 0.6[$), and \emph{defense} ($p(\text{defense}) \geq 0.6$).

\null 

At this point, we have a dataset with labels either through user reporting, or through prediction with high confidence thresholds. The next steps provide answers to our research questions.

\subsection{Measuring Stance Expression}

Our first research question is concerned with how people express their stance.
It is known that specific traits of language and linguistic focus are associated with abortion stances~\cite{zhang2016gender,sharma2017analyzing}. In terms of the Web, this is translated to hashtags and links. However, as new technologies enable other forms of communication, such as emojis~\cite{feng2019world}, we also ask ourselves whether these new ways of expression can be associated with stance.

A first glimpse into the potential answer of stance expression can be obtained by looking at the predictive power of features. Recall that gradient boosted trees are weak learners that are trained on a subset of the data, both in terms of observations and features. In XGB, the feature importance, defined as \emph{total gain}, quantifies the importance of a feature through the whole set of learners. There, an important feature not only helps in classification (tree building), but also in improving the accuracy of trees (boosting).

Therefore, we quantify how features relate to stance expression by grouping them into types (emojis, terms, links, {\it etc.}), and estimating whether the mean feature importance of each type is statistically different from the rest. We apply the Tukey's HSD test, which compares all possible pairs of means, and corrects for family error rates, effectively answering our research question.

\subsection{Inferring Stance Turnarounds}

Usually, the stances related to controversial issues are analyzed as a static phenomenon. Indeed, in our first question, we infer a single predominant stance for users in the whole period under analysis. However, a feature matrix can be built for any period of time, enabling the prediction of stance probabilities at different time windows within the dataset. As a result, for each time window, the difference in predicted probabilities may indicate whether users changed their leaning toward a stance or remained in their previous position. A change in stance is known as a \emph{turnaround}.

To analyze turnarounds, we compare two time windows separated by an event, in a similar way to how natural experiments work. For instance, legislation events and court decisions do not depend on the discussion in social media, but  they may stimulate such discussion (see Figure~\ref{fig:timeline}). 
We therefore select users who were active in both time windows, and build the corresponding feature matrices for each period. By applying the XGB stance classifier to each matrix, we obtain stance adherence probabilities for each user and period. The difference between two periods is defined as:
$$
\Delta(t_0, t_1, u) = p(\text{defense}_{t_1} \mid u) - p(\text{defense}_{t_0} \mid u),
$$
where $\Delta(t_0, t_1, u)$ lies in the range $[-1, 1]$. When $\Delta = 1$, the turnaround is completely toward \emph{defense}. When $\Delta = -1$, the turnaround is completely toward \emph{opposition} (recall that $p(\text{defense})$ is symmetrical with the probability of being in \emph{opposition}). A value of $\Delta = 0$ implies no change in opinion between periods for the corresponding user $u$.

To answer our second research question, we infer the relationship between profile characteristics and turnaround categories (or its absence, \emph{i.e.}, to \emph{remain} in a stance). As profile characteristics, we consider demographic features, as well as attributes easily obtained from each profile, including the number of followers, the number of followees, an activity ratio (number of tweets divided by account age in days), account age (in years), the stance predicted in the first period, and the usage of important terms in expressing opinion (a potential result from RQ1). This would allow to test social hypotheses, for instance, if popular accounts are less likely to change opinion due to their exposure, or whether as people get older, they become more conservative. 

We measure the relationship between the aforementioned variables and the value of $\Delta$ through adjusting a linear regression:
$$
\Delta = \beta_0 + \beta \cdot X + \epsilon,
$$
where $X$ is a user feature vector, $\beta$ is the regression coefficients vector, $\beta_0$ is the intercept, and $\epsilon$ is the error term. The value of each element in $\beta$ describes whether the corresponding feature is significant for stance turnaround. 

\null

The series of methods defined in this section, while drawing from previous work in the literature, provide a way to measure new aspects of stances in abortion as seen on micro-blogging platforms, namely, expressiveness and turnarounds. 
Next, we describe the results of applying these methods to our use case.

\section{Results}
\label{case_study}

We present the results of applying our methodology (Section~\ref{methods}) to the abortion debate in Argentina and Chile (Section~\ref{materials}). 

\begin{table}[t]
    \centering
    \footnotesize
    \caption{Classification metrics of demographic attributes and stance (5-fold cross validation, with the exception of the hold-out set for stance classification).}
    \label{table:classification_metrics}
    \begin{tabulary}{\linewidth}{lRll}
        \toprule
        Attribute &  Labeled N & Precision &  Recall \\
        \midrule
        Gender     & 212,302 & 0.71 $\pm$ 0.02 & 0.68 $\pm$ 0.03 \\
        Location   & 122,105 & 0.93 $\pm$ 0.04 & 0.93 $\pm$ 0.04 \\
        Age Cohort & 10,902 & 0.72 $\pm$ 0.01 & 0.64 $\pm$ 0.04 \\
        Stance & 26,994 & 0.93 $\pm$ 0.02 & 0.92 $\pm$ 0.03 \\
        Stance (Hold-out manually labeled) & 1,019 & 0.88 & 0.84 \\
        \bottomrule
        \end{tabulary}
    \end{table}
    
\begin{table}[t]
    \centering
    \footnotesize
    \caption{Distribution of users per demographic group.}
    \label{table:user_distribution}
\begin{tabulary}{\linewidth}{llRRRRR}
\toprule
Country & Gender & $<$ 18 &  18--29 &   30--39 &   $\geq$ 40 &  n/a \\
\midrule
Argentina & Female & 15.10 & 12.32 & 0.48 & 1.06 & 71.04 \\
      & Male & 5.48 & 8.98 & 0.87 & 2.77 & 81.91 \\ \midrule
Chile & Female & 1.75 & 7.68 & 4.67 & 4.14 & 81.76 \\
      & Male & 0.27 & 5.49 & 4.99 & 5.23 & 84.02 \\
\bottomrule
\end{tabulary}
\end{table}

We start with describing the performance of the demographic attributes (gender, location, age cohort) and stance classifiers applied to the 663,340 users in our dataset.
Table~\ref{table:classification_metrics} reports the performance metrics, estimated using 5-fold cross validation. We obtain the best performance for location and stance, precision of 0.93 for both and high recall (0.93 and 0.92, respectively). 
The performance of the gender prediction is below recent work using deep learning~\cite{wang2019demographic} by 0.18, and the 
performance of the age cohort prediction is above the same approach~\cite{wang2019demographic} by 0.23. 
Arguably the scenarios are not directly comparable, due to differences in using content-specific and general datasets. For instance, age prediction performance is good, but the size of the validation dataset is small. 

To account for potential prediction errors, we only consider predictions where the confidence is above specific thresholds. This particularly affects the age attribute, as 75.4\% of user age predictions were below the acceptance threshold (see Table~\ref{table:user_distribution} for the distribution of predicted demographic characteristics). We kept those users, as they may have posted relevant information. For location, Argentinian users makes 86\% of the dataset. An imbalance was expected, as Argentina has around three times the population of Chile. This pronounced imbalance suggests that the population in Argentina has different Twitter adoption patterns~\cite{graells2020representativeness}.

\begin{table}[t]
    \centering
    \footnotesize
    \caption{Top features of profile demographic characteristics.}
    \label{table:demographic_classification}
    \begin{tabulary}{\linewidth}{l|L}
        \toprule
        Attribute &  Top Features \\
        \midrule
        Gender     & \xelatexemoji{1f49a}, morir (to die), profile:ingeniero (male engineer), cuerpo (body), profile:n\_emojis, pibas (Argentinian word for girls),  ~name:\xelatexemoji{1f49a}, profile:\xelatexemoji{2665}, profile:abogada (female lawyer),\newline [sport words in the profile] \\ \midrule
        Location   &  timezone:Santiago, causales (grounds), \#aborto3causales (\#abortion3grounds), Chile, legal, \#abortolibre (\#freeabortion), libre (free), [social media keywords on the profile], favor, profile:n\_emojis \\ \midrule
        Age Cohort & dominio (domain), home page: instagram.com, objetores (objects), fav (social media slang), profile:n\_emojis, director, \#aborto3causales (\#abortion3grounds), profile:estudiante (student), cambiar (to change), Pichetto (Argentinian politician) \\
        \bottomrule
        \end{tabulary}
    \end{table}

Table~\ref{table:demographic_classification} shows the most important features in predicting demographic characteristics. Colored heart emojis play an important role in predicting gender (note that previous analysis of emoji usage only identified \xelatexemoji{2665} as associated with gender~\cite{chen2018through}), and words associated with the abortion debate are relevant to gender too (e.g~{\em morir} --to die--, and {\em cuerpo} --bo\-dy--). 
When predicting location, the most important features include timezone, local hashtags ({\it e.g.}, \emph{\#aborto3causales} refers to Chilean legislation), and the mentioning of other social-media sites on the biography, which may be related to different usage of social-media in the two  countries~\cite{graells2020representativeness}.
For age, important features include having a link to an Instagram account, using social-media slang ({\it e.g.}, \emph{fav}), the total number of emojis in the description, and having the word {\it estudiante} --student-- in the biography. 
These observations are in general not surprising, but they show insights on the usage of emojis, in terms of which ones (gender) and how many (age) help in classifying users.

\begin{figure}[t]
\centering
\includegraphics[width=0.9\linewidth]{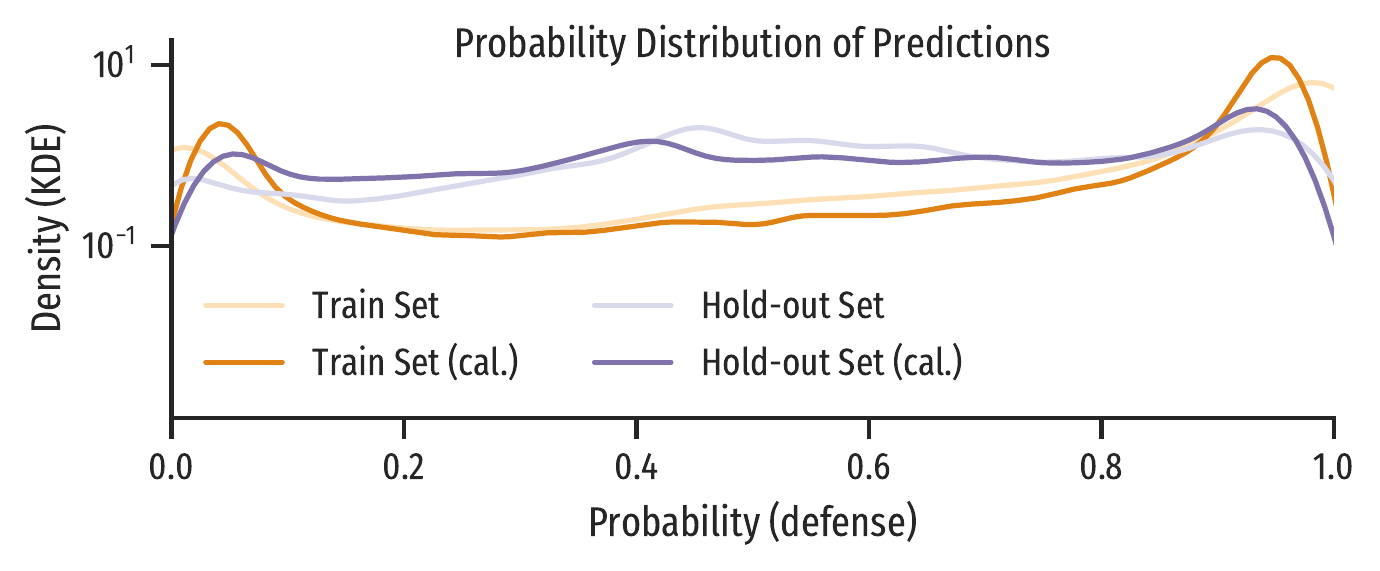}
%\vspace*{-0.3cm}
\caption{Probability distribution of predictions for the training set (self-reported profiles) and the hold-out set used for calibration.}
\label{fig:stance_probabilities}
%\vspace*{-0.3cm}
\end{figure}

\begin{figure}[t]
\centering
\includegraphics[width=\linewidth]{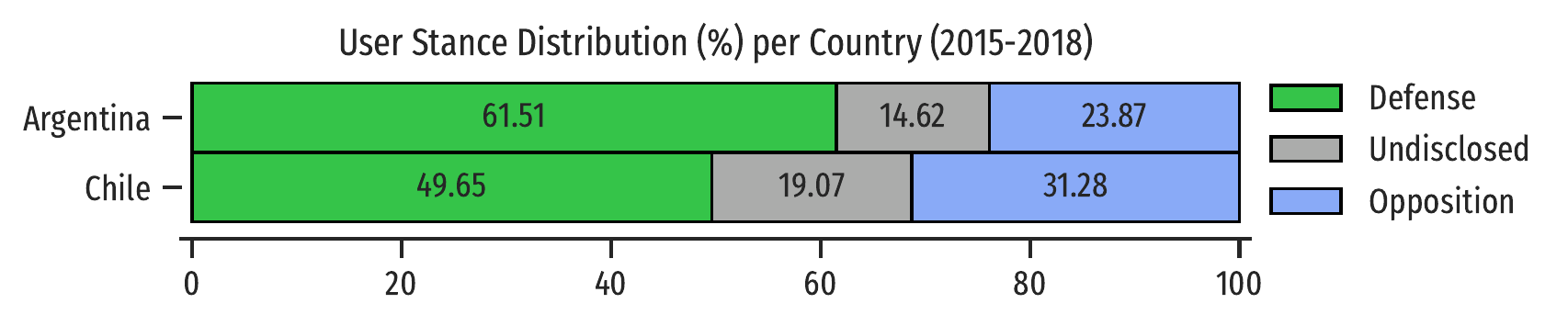}
%\vspace*{-0.3cm}
\caption{Stance distribution per country.}
\label{fig:stance_per_country}
%\vspace*{-0.2cm}
\end{figure}

With respect to stance prediction, the performance exceeds typical values, where precision lies around 85\%~\cite{cohen2013classifying}.  
In the labeled set, all users had a self-reported stance, which implied an explicit position on abortion. However, the performance achieved on the hold-out set is slightly worse (0.88 precision, 0.84 recall), but is still within a good performance range. This set was comprised with 1,019 manually labeled users, randomly selected from the unlabeled group. Its proportion of labels is slightly more balanced than in the self-reported set: 80\% \emph{defense}, 20\% \emph{opposition}; compared with 85\% and 15\%, respectively. Figure~\ref{fig:stance_probabilities} shows the distribution of calibrated probability outcomes from the classifier on both datasets, automatically labeled (training set) and hold-out (manually labeled), where calibration was performed using Platt scaling. The training set exhibits two probability peaks, one at each extreme, whereas the hold-out set does not exhibit this behavior, and shows a third peak around $0.4$. If we consider an \emph{undisclosed} category, this could signal that people who do not express explicitly their position are more likely to be conservative (\emph{opposition}). To account for this uncertainty, we considered the \emph{undisclosed} category in our analyses. In total, 59.86\% of users are in \emph{defense}, 15.24\% are in \emph{undisclosed}, and 24.90\% are in \emph{opposition} (see Figure~\ref{fig:stance_per_country} for the distribution per country).

In summary, the prediction of profile characteristics yielded results that are on-par with the state-of-the-art in terms of stance and location, which we need to answer RQ1. To answer RQ2, we also need gender and age, so in the regression analysis we only considered profiles with predictions generated with high confidence.

\begin{table*}[t]
    \centering
    \footnotesize
    \caption{Top-30 features for stance classification weighted by association to each country.}
    \label{table:country_stance_features}
\begin{tabulary}{\linewidth}{lLLL}
\toprule
Country &  Defense &  Opposition & Undisclosed \\
\midrule
Argentina &  \xelatexemoji{1f49a}, legal, clandestino (clandestine), seguro (safe), gratuito (free), sexual, morir (to die), decidir (to decide), anticonceptivos (contraceptive), educación (education), abortar (to abort), clandestinidad (clandestinity), mujeres (women), años (years), pibas (girls) &  \xelatexemoji{1f499}, matar (to kill), \#aborto (\#abortion), profile:n\_emojis, vida (life), solución (solution), @mauriciomacri, Macri (Argentinian President), nacer (to born), humano (human), niño (kid), @marianoobarrio, abortistas (abortists), asesinato (murder), muerte (death) &  profile:n\_emojis, [profession words in the profile], [social media words in the profile], [education words in the profile], [family words in the profile], tema (topic), católico (catholic), profile:siempre (always), profile:\xelatexemoji{1f499}, timezone:Santiago, aborto (abortion), profile:periodista (journalist), ateo (atheist), Cristina (former Argentinian President), favor (favor) \\ \midrule
Chile     &  mujeres (women), legal, causales (grounds), \xelatexemoji{1f49a}, decidir (to decide), tres (three), despenalización (decriminalization), seguro (safe), gratuito (free), derechos (rights), UDI (far-right Chilean political party), años (years), clandestino (clandestine), protocolo (protocol), bit.ly (link sharing URL) &  vida (life), \xelatexemoji{1f499}, muerte (death), izquierda (left), niños (kids), asesinato (murder), matar (to kill), ideología (ideology), nacer (to born), negocio (biz), humano (human), aborto (abortion), niño (kid), inocentes (innocent), crimen (crime) &  timezone:Santiago, profile:n\_emojis, [profession words in the profile], [education words in the profile], estaría (would be), [family words in the profile], profile:música (music), profile:periodista (journalist), profile:siempre (always), provida (profile), profile:gusta (like), [social media words in the profile], profile:fanático (male fan of), gente (people), profile:ex \\
\bottomrule
\end{tabulary}
\end{table*}

\begin{figure}[t]
\centering
\includegraphics[width=0.95\linewidth]{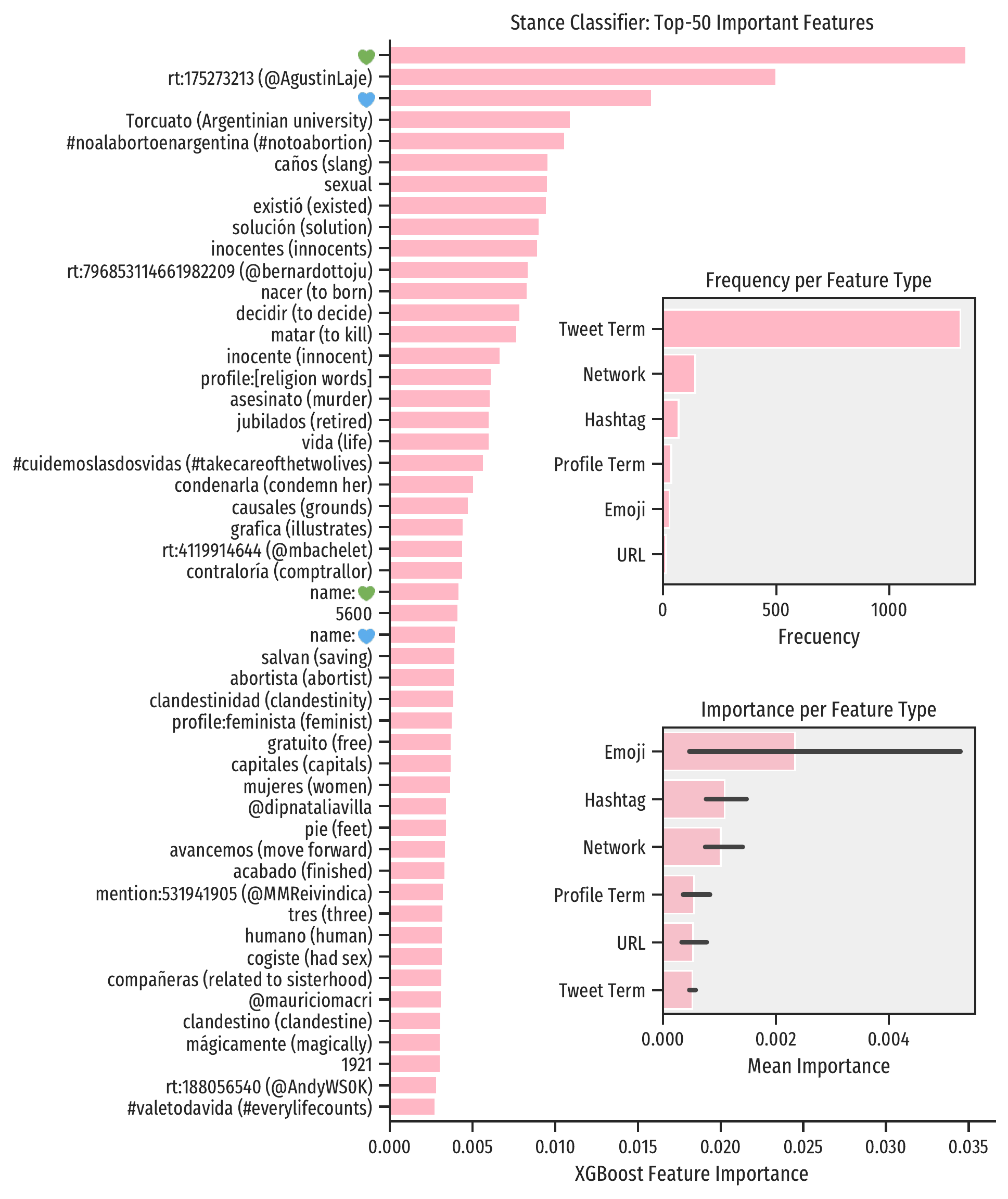}
%\vspace*{-0.3cm}
\caption{Relevant features for stance prediction. The larger bar chart includes the top-50 features, the embedded bar charts show the frequency of each feature type, and each type's mean and standard deviation of the corresponding importance.}
\label{fig:stance_feature_importance}
%\vspace*{-0.5cm}
\end{figure}

\subsection{Stance Expressiveness}

Our first research question is: \emph{how people make use of new technologies to express their positions on controversial issues?} In other words, we want to understand how new forms of communication on the Web enable both, prediction of stance, but also expression, in the sense of explicit adherence to a stance.

Figure~\ref{fig:stance_feature_importance} shows the top-50 most important features for stance classification, from a total of 1,602 features. Two different emojis appear on the top-50 (\xelatexemoji{1f49a} \& \xelatexemoji{1f499}), in four different features, as they can appear as terms in tweet content, and as terms in a profile's name.
Since feature importance alone does not indicate association with a stance, Table~\ref{table:country_stance_features} shows the top-30 features associated with each country and stance (including \emph{undisclosed}). The table confirms that the green heart is associated with \emph{defense}, whereas the blue heart is associated with \emph{opposition}, which are the colors adopted by activists to symbolize their position on abortion stance in physical manifestations in Latin America~\cite{sutton2020abortion}. Other terms show expected associations, such as a focus on \emph{women} (mujeres) and the \emph{right to choose} (decidir) in \emph{defense}, or the focus on \emph{life/death} (vida/muerte/matar), and the \emph{right to be born} (nacer) in \emph{opposition}. Features associated with \emph{undisclosed} are mixed, with a diverse set of generic profile features.

Figure~\ref{fig:stance_feature_importance} contains two additional charts, with frequency and mean importance per type of feature: emojis, network features (interactions), hashtags, terms in tweets and profile biography, and URLs. The frequency of emojis is smaller than most of the other features (29), with exception of URLs. However, on average, it is one of the most important (mean XGB importance: 0.00234). 
To validate whether there are differences in the main features, we applied the multiple comparison Tukey's HSD test.
The differences between emojis and the rest of the feature types are all significant (rejecting the null hypothesis of no differences, with $p < 0.001$), while the difference between other types is not statistically different, with the exception of tweet terms with network interactions and hashtags. 
Hence, even though few emojis are related to stance, their predictive power is higher than language associated with abortion stances, and higher than other types of features.

As answer to RQ1, we observe that emojis significantly express stance in a way that mirrors physical manifestations, either as a symbol of adherence (in the name), or as a symbol of support in discussion (in tweet content). We discuss this further in Section~\ref{discussion}.

\begin{figure}[t]
\centering
\includegraphics[width=0.85\linewidth]{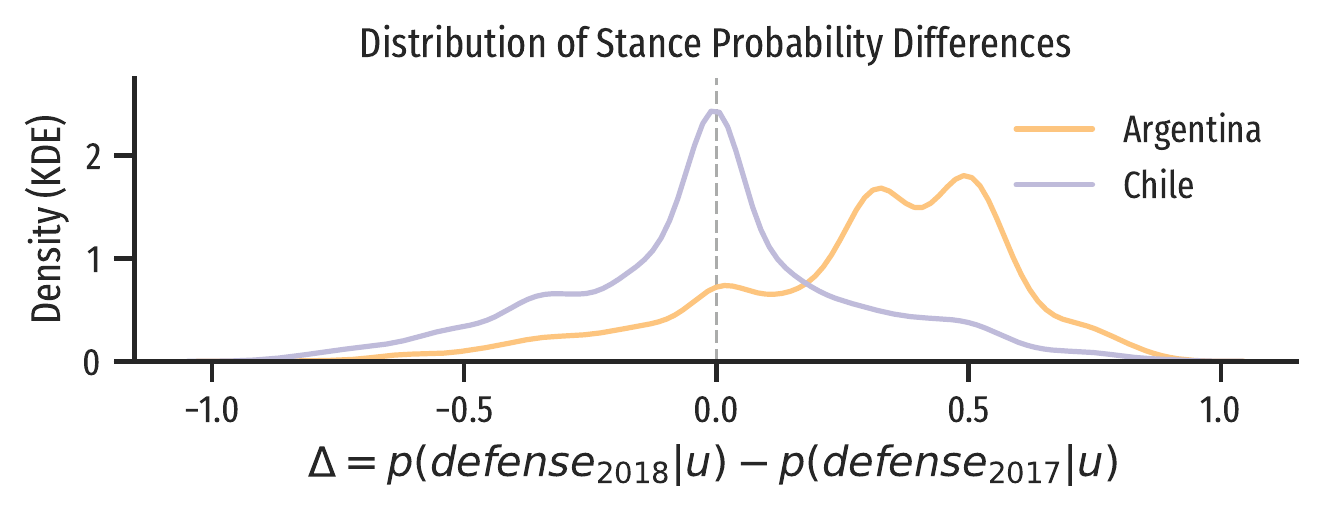}
\caption{Probability distribution of turnaround.}
\label{fig:turnaround_probabilities}
\end{figure}

\subsection{Stance Turnarounds}

Our second research question is: \emph{which demographic and profile factors characterize opinion change?}
To answer it, we study two time windows of three months in two consecutive years: May, June, and July, in 2017 and 2018.
During these months in 2017, the Chilean Senate approved the abortion bill proposed by President Bachelet (see Figure~\ref{fig:timeline}). One year later, a free abortion bill was being discussed in Argentina, during the government of President Macri. Given that the new Chilean law allowed abortion on three grounds, the possible approval of a free abortion in Argentina was likely to influence the following discussion in Chile. 
Thus, we hypothesize that the discussion in Argentina during 2018 was a moment to express a potentially updated view on abortion, with respect to the previous year, particularly for Chileans, but also for Argentinians who participated in the 2017 debate.

\begin{figure}[t]
\centering
\includegraphics[width=0.95\linewidth]{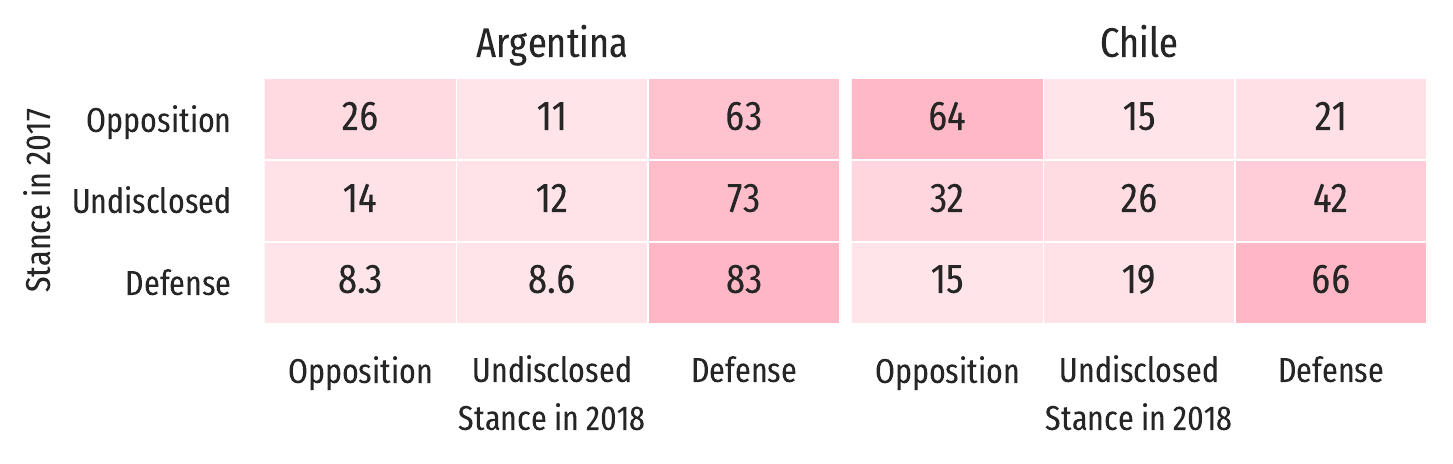}
\caption{Stance turnarounds per country.}
\label{fig:country_turnarounds}
\end{figure}

In our dataset, 11,931 users participated in the discussion during both periods, and for whom we have a predicted age and gender with high confidence.
In this subset, the proportion of Chilean accounts is greater than in the full dataset (31\% instead of 14\%), as Argentinian users tended to participate in the discussion only in 2018. 
Figure~\ref{fig:turnaround_probabilities} shows the distribution of probability differences for the users appearing in both periods. Chilean users present a symmetric distribution, suggesting that the majority of users maintained their views (see Figure~\ref{fig:country_turnarounds}), whereas Argentinian users exhibit a clear tendency to change \emph{toward defense}.

\begin{figure}[t]
\centering
\includegraphics[width=\linewidth]{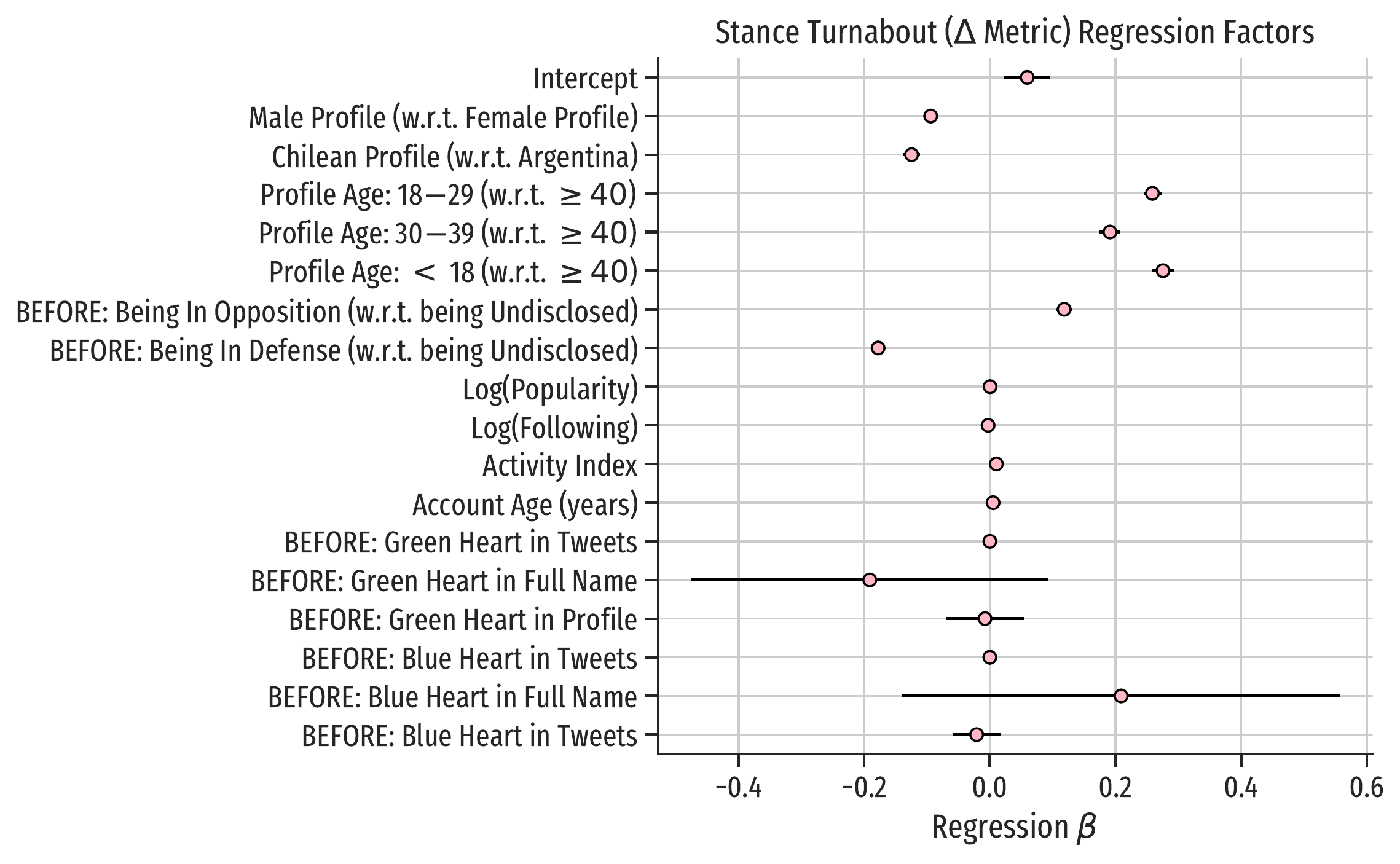} 
%\vspace*{-0.5cm}
\caption{Regression factors and their confidence intervals from stance turnaround regression.}
\label{fig:turnabout_regression}
%\vspace*{-0.5cm}
\end{figure}

To infer which demographic and content factors were important in determining a turnaround, we performed a linear regression over the turnaround metric $\Delta$. 
The independent variables of the regression included the demographic factors predicted for each user and profile characteristics, such as number of followers, followees, activity ratio, account age, and the usage of colored emojis. The number of followers, followees and activity ratio were log-transformed due to their skewed distributions. 

The model exhibited a good fit of the data (adjusted $R^2 =$ 0.397, $MSE =$ 0.1011, $F =$ 524.2, $p < 0.001$, Log-Likelihood $LL =$ -458.70). 
According to the adjusted $R^2$ coefficient, the model explains 40\% of the variance in stance turnaround. 
Figure~\ref{fig:turnabout_regression} shows the regression factors with their confidence intervals.
As some variables were categorical, the regression used dummy variable coding, and thus, some categories are reference values that are encoded in the intercept.
Keeping in mind that positive factors indicate leaning toward \emph{defense}, and negative factors indicate leaning toward \emph{opposition}, one can see that males tend to turnaround toward \emph{opposition} (male $\beta =$ -0.0942) in comparison to females. Also, as people gets older, their turnaround difference changes from toward \emph{defense} to \emph{opposition} (from $\beta =$ 0.2757 for $<$ 18, $\beta =$ 0.2587 for 18--29, and $\beta =$ 0.1912 for 30--39, with $\geq$ 40 as reference). 

Even though the distribution of differences in Chile is symmetrical, in comparison to Argentina, Chileans lean toward \emph{opposition} ($\beta =$ -0.1624). Network metrics (popularity and following) are not significant; conversely, the amount of activity as well as the age of the account are positively related with turnaround toward \emph{defense} ($\beta =$ 0.0103 and $\beta =$ 0.0051, respectively), meaning that active and old accounts are more likely to be in \emph{defense}.
Surprisingly, the usage of emojis was only significant for \xelatexemoji{1f49a} in tweet content during the first period ($\beta =$ 0.0051). 

To conclude, stance turnaround happens. Moreover, publicly available features (profile characteristics and emoji usage) plus other predictable attributes (demographics) provide a solid baseline to analyze stance turnaround in quantitative terms, effectively answering our research question. 

\section{Discussion}
\label{discussion}

We discuss the main implications of this work, in terms of \emph{technological means of expression} and \emph{stance turnarounds and demographic composition}. We also discuss the limitations of our work.
%\vspace{0.1cm}

\paragraph{Technological Means of Expression}
The identification of political stance through color is not new. Women's causes in the last century have been associated with green, white, and purple, expressed through clothing accessories~\cite{sawer2007wearing}. 
However, in non-Spanish speaking countries these causes did not include abortion rights. In Latin America, the usage of colors also emerged from feminist movements~\cite{sutton2020abortion}, and, as our results indicate, this phenomenon is manifested in web platforms, not only for feminist positions regarding abortion (\xelatexemoji{1f49a}), but also for opposition to abortion rights (\xelatexemoji{1f499}). 
In the last decade, emoji input has been a common feature in modern user interfaces, although they have been commonly used as a emotional device rather than a political one. Compared to other web-based means of expression, emojis have emerged in the abortion discussion using out-of-domain symbolism. For instance, hashtags tend to include concepts, dates, and mottos relevant to the issue under discussion, and are indeed related with protests~\cite{steinert2015online}. Conversely, these emojis have an arguably neutral appearance (a metaphorical heart) and a color that is not necessarily associated with a political party~\cite{sutton2020abortion}.
Their current prevalence, and the adoption of younger generations of emojis as a natural device to engage in conversation, imply that future studies should reconsider whether text-only insights still apply, and whether removing non-domain/non-textual context eliminates powerful signals.
%\vspace{0.1cm}

\paragraph{Stance, Turnarounds and Demographic Composition}
Abortion is also a public health issue, and domain experts in health have pointed out that, even though the Twitter population is biased~\cite{RBY18}, if done carefully, Twitter allows to measure what ``the public is \emph{actually} seeing,'' an understanding that would help to design communication strategies regarding abortion legislation and medical practice~\cite{han2017tweeting}. Previous work has improved the understanding of \emph{who} is this public, by measuring its representativeness of the general population~\cite{graells2020representativeness}; here we observed that demographic factors and profile characteristics explain a significant amount of the variance in stance turnaround between two specific periods. A main implication of this result is that the proposed methodology can be applied to measure how people react to specific events, which may be political, but may also be communication strategies to improve views on public health interventions. 
%\vspace{0.1cm}

\paragraph{Limitations}
Our work is scoped in several aspects. 
First, the lack of ground truth makes our evaluation strictly based on the data labeled or inferred from self-reported information. Moreover, as the Twitter Streaming API is a sample, the dataset does not cover the entire discussion on the platform. However, our previous work showed that insights derived from this dataset match those from nationally representative surveys~\cite{graells2019representative}.
Second, our methodology needs further validation. Individual steps can be evaluated, such as a formal comparison of the classifier with other approaches~\cite{wang2019demographic}, and measuring the classifier performance in turnaround scenarios with labeled data. 
Finally, our main insight is the identification of previously unseen features, but we cannot rule out the possibility that other important features were excluded due to our seed keyword identification processes. A sensitivity analysis of seed keywords should be performed.

\section{Conclusions}
\label{conclusions}

% aims 
In this paper we characterized the expressiveness and turnarounds of stances in controversial issues. 
We did so by performing a longitudinal analysis of the abortion debate in two Latin-American countries, Argentina and Chile.
% results
We empirically quantified these phenomena using established methods from the literature, finding that, even though abortion is not a new topic, the resources being employed to take a stance and express it are evolving, as the Web does;
and that it is possible to characterize changes in opinion according to demographic factors and profile characteristics.
% techniques with good results
Gradient-boosted trees showed potential in the context of abortion debate, enabling the propagation of self-reported attributes and our own knowledge about the issue, but also the identification of novel ways in which people express their stance. The inclusion of demographic and profile characteristics on turnaround analysis show that it is possible to evaluate the effect on stance of legislative events, and potentially other types of interventions, such as communication strategies for public health awareness.

% future work, long-term impact of the work
In 2010, in the rise of Twitter and other social platforms, a famous article from The New Yorker stated that the revolution will not be tweeted~\cite{gladwell2010revolution}. 
Times have changed: our results show that these platforms reflect important aspects of physical manifestation in events that span multiple years, and that people adopt new means to express themselves and communicate their positions.
The Web has the potential to act as a positive influence on how people relate to others, and future work should look into how new \& old ways of communication (such as emojis, memes, and short videos) are related to people's political beliefs, and how people of opposing views could be brought together through them.

%\vspace{0.15cm}

%\begin{acks}
\paragraph{Acknowledgments} We thank the authors of the twemoji package (CC-BY4), Ana Freire for insightful discussion, and the anonymous reviewers for their insightful comments.
%\end{acks}

\bibliographystyle{ACM-Reference-Format}
\bibliography{main}

\end{document}